# Reducing the Energy Cost of Inference via In-sensor Information Processing

**Sai Zhang Mingu Kang Charbel Sakr Naresh Shanbhag**
Department of Electrical and Computer Engineering
University of Illinois at Urbana-Champaign
Urbana, IL 61801
{szhang12, mkang17, sakr2,shanbhag}@illinois.edu

## Abstract

There is much interest in incorporating inference capabilities into sensor-rich embedded platforms such as autonomous vehicles, wearables, and others. A central problem in the design of such systems is the need to extract information locally from sensed data on a severely limited energy budget. This necessitates the design of energy-efficient realizations of machine learning systems for processing sensory data - sensory embedded system. A typical sensory embedded system enforces a physical separation between sensing and computational subsystems - a separation mandated by the differing requirements of the sensing and computational functions. Sensing being done on a noisy analog statistical fabric while computing being executed von Neumann style on a deterministic digital fabric. As a consequence, the energy consumption in such systems tends to be dominated by the energy consumed in transferring data over the sensor-processor interface (communication energy) and the energy consumed in processing the data in digital processor (computational energy). In this article, we propose an in-sensor computing architecture which (mostly) eliminates the sensor-processor interface by embedding inference computations in the noisy sensor fabric in analog and retraining the hyperparameters in order to compensate for non-ideal computations. The resulting architecture referred to as the Compute Sensor - a sensor that computes in addition to sensing - represents a radical departure from the conventional. We show that a Compute Sensor for image data can be designed by embedding both feature extraction and classification functions in the analog domain in close proximity to the CMOS active pixel sensor (APS) array. Significant gains in energy-efficiency are demonstrated using behavioral and energy models in a commercial semiconductor process technology. In the process, the Compute Sensor creates a unique opportunity to develop machine learning algorithms for information extraction from data on a noisy underlying computational fabric.

## 1 Introduction

Computing platforms that exhibit inference and decision-making capabilities are much valued today. Incorporating these capabilities requires the implementation of complex machine learning algorithms and large data sets on which to operate on. However, due to their high computational and storage complexity, today machine learning systems are deployed in the cloud and on large-scale general-purpose computing platforms such as CPU and GPU-based clusters [5]. This "*intelligence in the cloud*" architecture presents a challenge for untethered (embedded) platforms such as cell phones, autonomous unmanned vehicles, and wearables. For such embedded platforms to exhibit inference capabilities, it requires the transmission of voluminous quantities of locally acquired raw sensor data to the cloud. The resulting energy and latency costs are significant. Indeed, recent projections [1] indicate that the traffic to the cloud consumes $9\times$ more energy compared to that in the data center itself.

Therefore, there is much interest in implementing machine learning kernels on embedded computing platforms. Such platforms, however, have stringent limits on available energy, computation, and storage resources. This necessitates a fresh look at the design of learning algorithms, architectures and integrated circuits (ICs) realizations. Initial work in this direction has already begun with recent IC prototypes of inference systems [7, 14, 3, 20].

Figure 1(a) shows a conventional architecture of an embedded vision system. Image data is first acquired via an $M_r$ row $\times$ $M_c$ column active pixel sensor (APS) array whose analog pixel values are sensed sequentially in a row-wise fashion, and then converted into digital samples by the sample-and-hold (S/H) and the analog-to-digital converter (ADC), and then streamed out by the read-out (RD) circuitry to a back-end digital processor which implements feature extraction and classification function to obtain the final decision $\hat{y}$. A digital trainer block computes the hyperparameters in supervised learning mode. This physical separation between sensing and processing subsystems is made unavoidable because sensing is intrinsically an analog process (photon energy is converted into continuous-valued electric current) while information processing is intrinsically digital. The energy dissipation in such a system is dominated by two sources:

- The energy required to move the data over the sensor-processor interface comprising the ADC, RD and the interconnect to the digital processor, i.e., *communication energy*, and
- the energy consumed in processing the data using digital circuits which by nature are high signal-to-noise ratio (SNR), i.e., *computational energy*.

We assume that the storage energy is negligible in the streaming architecture being considered here. We employed energy data from [4] for a CMOS image sensor consisting of a $32 \times 32$ APS array and the associated interface circuits, and estimated the computational energy needed to implement a principal component analysis (PCA) [13] engine and a support vector machine (SVM) [22] in a $65\,\mathrm{nm}$ CMOS process operating at a throughput of 32 frames/s. This analysis indicates that the communication and computational energies are approximately $53\%$ and $41\%$, respectively, for a combined total of $94\%$. An impactful solution to the energy problem needs to reduce both components of energy - communication and computational energy.

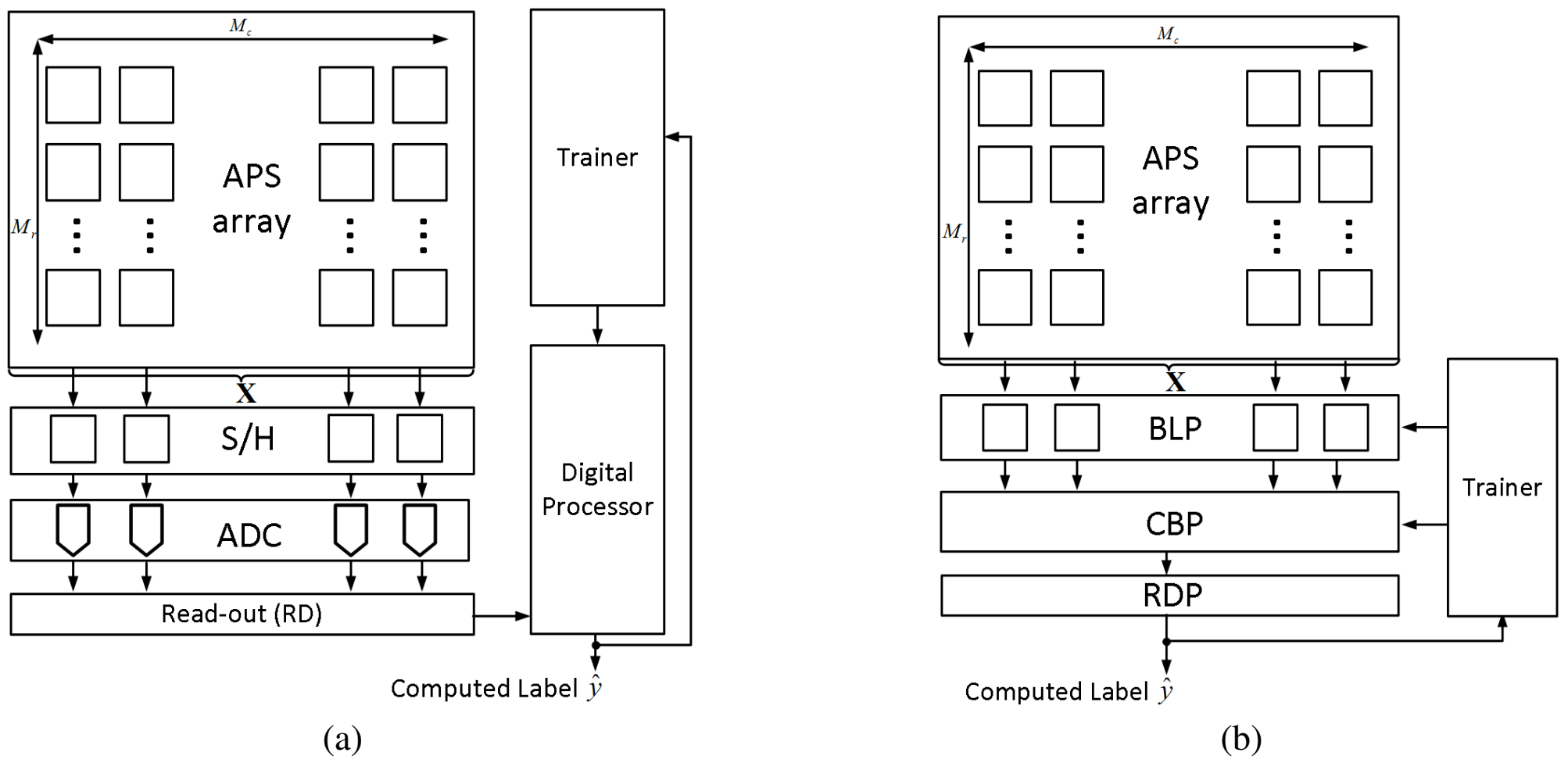

Figure 1: A typical embedded vision platform: (a) conventional architecture, and (b) the proposed Compute Sensor architecture.

In this paper, we propose the Compute Sensor shown in Figure 1(b) - a sensory system that senses *and* processes the sensed data thereby integrating both data acquisition and information extraction functionalities. The Compute Sensor architecture consists of a data processing engine and a training engine. The data processing engine is a cascade of: 1) the APS array which is identical to the conventional architecture, 2) a bit-line processor (BLP) whose physical dimensions are matched to that of the APS array in order to perform pixel-wise operation such as sample-and-hold (S/H), scaling, and absolute difference but no ADC, 3) a cross bit-line processor (CBP) to perform data dimensionality reduction operations such as dot product, filtering, sum-of-absolute difference (SAD), mean square, followed by an ADC that operates on the reduced dimensionality data and feeds it into 4) the residual digital processor (RDP) which implements very simple digital computations needed to obtain the final decision $\hat{y}$. Unlike the conventional architecture, both the BLP and CBP in the Compute Sensor operate in the analog domain. The trainer is digital.

The Compute Sensor eliminates both the traditional sensor-processor interface, and the high-SNR/high-energy digital processing by moving feature extraction and classification functions into the analog domain in close proximity to the APS array. The Compute Sensor leverages the intrinsic ability of machine learning algorithms to extract information from noisy and often incomplete data to provide robust inference in presence of non-ideal computations. We demonstrate a Compute Sensor that incorporates a PCA-based feature extractor and a support vector machine (SVM). Using circuit characterized behavioral and energy models in a $65\,\mathrm{nm}$ CMOS process, we show that the Compute Sensor is able to achieve a detection accuracy greater than $94.7\%$ using the Caltech101 dataset [8], which is within $0.5\%$ of that achieved by an ideal floating point algorithm. Furthermore, the Compute Sensor is able to compensate for variations in the electrical parameters of the transistors in the APS array caused by finite tolerances of the semiconductor manufacturing process by retraining in presence of these non-idealities. As a result the Compute Sensor consumes $7\times$ to $17\times$ lower energy than the conventional architecture for the same level of accuracy. Thus, this paper highlights the potential for conducting algorithmic research that accounts for platform resource-constraints such as energy, storage, and computation.

The rest of the paper is organized as follows. Section 2 presents the necessary background and establishes notation. Section 3 presents the Compute Sensor architecture incorporating PCA and SVM algorithms, and the behavioral and energy models in a $65\,\mathrm{nm}$ CMOS processes. Simulation results are shown in Section 4, and discussions are provided in Section 5.

## 2 Background

### 2.1 Relevant Work

Previous work in integrating computation into a CMOS image sensor array falls into two categories. In the first, the pixel architecture is modified to enable computations such as 2D convolution [18], image filtering [6, 16], compressive image sensing [21], matrix transformations [2], Gaussian pyramid [9], and image decomposition [15]. These approaches suffer from a loss in *fill-factor* or the spatial resolution because each modified pixel occupies an area than can be as high as $8\times$ higher than the standard pixel architecture. This is why in designing the Compute Sensor, we choose to leave the APS array intact. In the second, very simple analog processing functions are embedded in the periphery of the APS array. These include convolution [10], random projections for compressed sensing [19], difference of Gaussian (DOG) [17]. None of these approaches exploit the ability of learning algorithms to adapt to circuit non-idealities that arise when computation is implemented in low-SNR analog sensing fabrics. Retraining a digital processor implementing a feature extractor [23] has shown that the retrained hyperparameters can compensate for computational errors such as those caused by defects. We employ retraining to compensate for non-idealities that are arise when computation is attempted in the analog domain.

### 2.2 Principle Component Analysis (PCA)

PCA is a widely used method for dimensionality reduction. This reduction is accomplished by projecting the data vector $\mathbf{x}_n \in \mathcal{R}^M$ ($n = 1, ..., N$, is the sample index and $N$ is the total number of samples in the dataset) onto a set of orthonormal principal components $\boldsymbol{\alpha}_k \in \mathcal{R}^M, k = [1, ..., K]$. These principal components are the top $K$ variance maximizing eigenvectors of the sample covariance matrix $\sum_{n=1}^{N} \mathbf{x}_n \mathbf{x}_n^T$ [13]. Hence, the reduced dimension (feature) vectors $\mathbf{f} \in \mathcal{R}^K$ are obtained as:

$$\mathbf{f} = \mathbf{A}\mathbf{x} \tag{1}$$

where $\mathbf{A} = [\boldsymbol{\alpha}_1, ..., \boldsymbol{\alpha}_K]^T \in \mathcal{R}^{K\times M}$ is the eigenmatrix, and $\mathbf{x} \in \mathcal{R}^M$ is the *test* data vector. In the CMOS image sensor shown in Figure 1(a), $\mathbf{x}$ is obtained from the APS array and $M = M_r M_c$, where $M_r$ and $M_c$ are the number of rows and number of columns, respectively, in the APS array.

### 2.3 Support Vector Machine (SVM)

The SVM [22] is a popular supervised learning method for classification and regression. In SVM, the trained model is represented by:

$$y_o = \mathbf{w}_s^T \mathbf{f} - b \tag{2}$$

where $\mathbf{w}_s \in \mathcal{R}^K$ is the optimum weight vector, and $\mathbf{f} \in \mathcal{R}^K$ is the *test* feature vector. It can be shown that the optimum weight vector $\mathbf{w}_s$ can be described in terms of feature vectors that lie on the margins, i.e., support vectors:

$$\mathbf{w}_s = \sum_{n=1}^{N_s} \beta_n y_n \mathbf{f}_{s,n} \tag{3}$$

where $y_n$, $N_s$ and $\mathbf{f}_{s,n}$ are the label, the number of support vectors, and the $n^{th}$ support vector, respectively. The SVM's classification accuracy is denoted by $p_c = Pr\{\hat{y} = y\}$, where $\hat{y} = sgn(y_o)$ is the computed label and $y$ is the true label.

## 3 The Compute Sensor

This section presents the proposed Compute Sensor architecture for implementing the PCA and SVM, along with architectural level functional and energy models in a $65\,\mathrm{nm}$ CMOS process. These models are employed to study the effectiveness of retraining on compensating for analog non-idealities, and for estimating the energy consumption.

### 3.1 Architecture

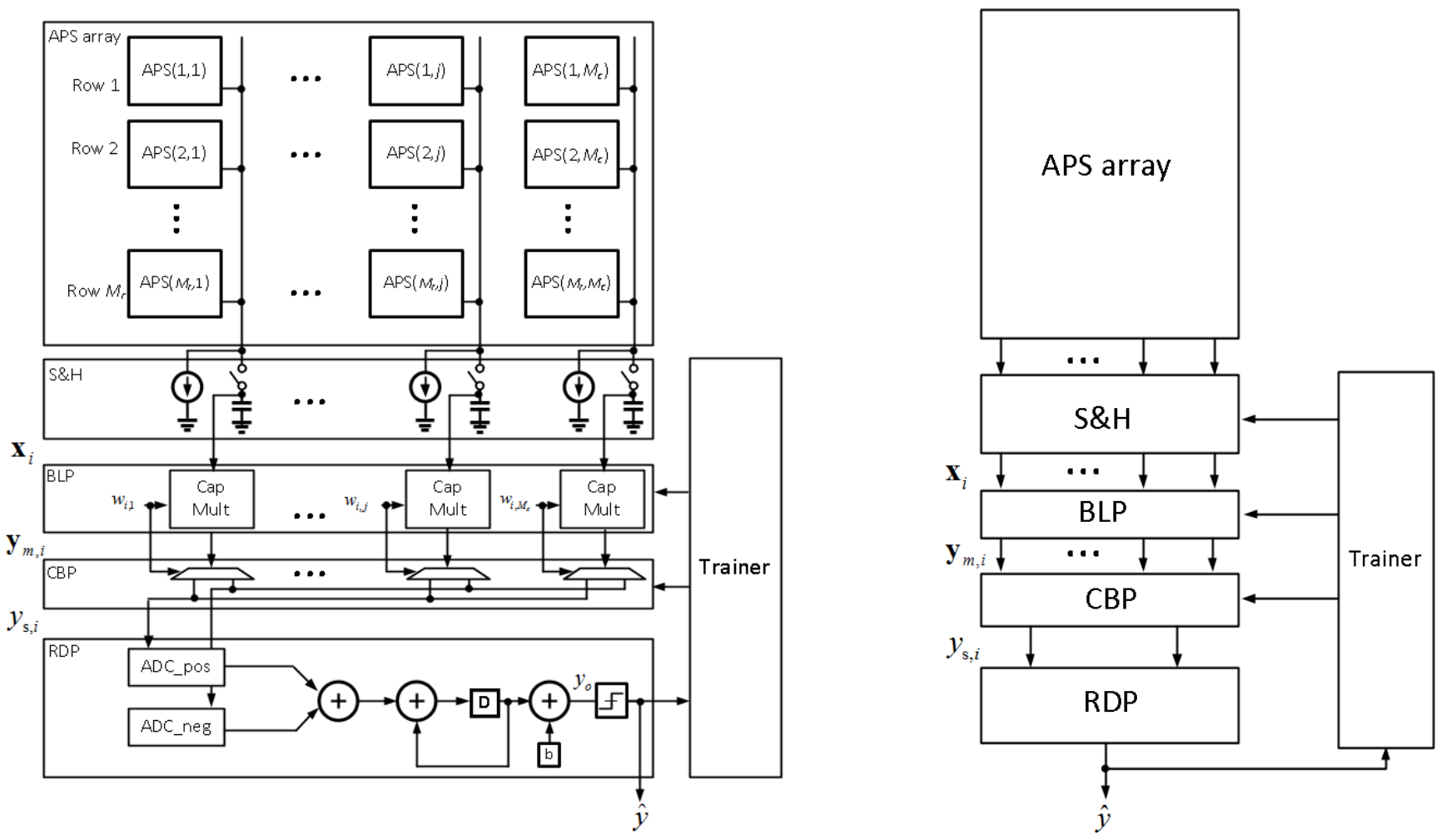

(a) (b)

Figure 2: Compute Sensor implementing PCA and SVM: a) the architecture, and b) the behavioral model.

The general Compute Sensor architecture in Figure 1(b) enforces a specific sequence of functions - acquire data in the APS array of size $M = M_r M_c$, bit-line processing, cross bit-line processing, followed by residual digital processing. Bit-line processing involves scalar operations while the cross bit-line processing results in dimensionality reduction. For example, bit-line operations could be the product of scalar data values and scalar weights, while cross bit-line processing would sum up these scalar products to generate a dot product. In the following, we remember that $M = M_c M_r$. Keeping these architectural constraint in mind, we exploit the linearity of the PCA and SVM computations in (1) and (2) to combine them as follows:

$$y_o = \mathbf{w}_s^T \mathbf{A}\mathbf{x} - b = \mathbf{w}^T \mathbf{x} - b \tag{4}$$

where $\mathbf{w}^T = \mathbf{w}_s^T \mathbf{A} = \left[\mathbf{w}_1^T, \ldots, \mathbf{w}_{M_r}^T\right] \in \mathcal{R}^{1\times M}$, $\mathbf{w}_i \in \mathcal{R}^{M_c}$, and $\mathbf{x} \in \mathcal{R}^M$, where $\mathbf{x}^T = \left[\mathbf{x}_1^T, \ldots, \mathbf{x}_{M_r}^T\right]$ and $\mathbf{x}_i \in \mathcal{R}^{M_c}$. The composite weight vector $\mathbf{w}$ can be obtained directly via SVM training methods. The data acquisition in the APS array occurs sequentially in a row-by-row fashion. In order to accomodate this constraint, we rewrite (4) as follows:

$$y_o = \left[\mathbf{w}_1^T, \ldots, \mathbf{w}_{M_r}^T\right] \begin{bmatrix} \mathbf{x}_1 \\ \vdots \\ \mathbf{x}_{M_r} \end{bmatrix} - b = \sum_{i=1}^{M_r} \mathbf{w}_i^T \mathbf{x}_i - b \quad (5)$$

This simple step enables us to implement multiplications operations involved in computing the dot product $\mathbf{w}_i^T \mathbf{x}_i$ (5) in the BLP consisting of an array of $M_c$ capacitive multipliers, and the addition of these products in a charge sharing-based adder in the CBP. The Compute Sensor's classification accuracy $p_c = Pr\{\hat{y} = y\}$ is calculated in the same manner as that in the conventional system.

### 3.2 Behavioral and Energy Models

Behavioral models describe the input-output relationship of the various blocks constituting the Compute Sensor while accounting for circuit non-idealities. These models can be employed in system simulations to estimate the performance of algorithms implemented on the Compute Sensor. Figure 2(b) shows that the first two stages of the Compute Sensor - the APS array and the S/H blocks - maps light energy incident on the $i^{th}$ row of pixels to $\mathbf{x}_i$ as follows:

$$\mathbf{x}_i = x_{max}\mathbf{1} - \gamma \mathbf{I}_i + \boldsymbol{\eta}_{s,i} + \boldsymbol{\eta}_{a,i} \quad (6)$$

where $\mathbf{x}_i \in \mathcal{R}^{M_c}$ is a discrete-time continuous-amplitude voltage representation of the luminous exposure $\mathbf{I}_i$ incident on the $i^{th}$ row of pixels, $x_{max}$ is the maximum output, $\mathbf{1}$ is a column vector with all ones, $\gamma$ is the conversion gain, $\boldsymbol{\eta}_{s,i} \in \mathcal{R}^{M_c}$ is a vector of samples from $\mathcal{N}(0, \sigma_s)$ representing the impact of spatial mismatch in device parameters across the APS array, and $\boldsymbol{\eta}_{a,i} \in \mathcal{R}^{M_c}$ is a vector of samples from $\sim \mathcal{N}(0, \sigma_a)$ representing the thermal noise in the APS array.

The BLP scales each pixel value by the weight $w_{i,j}$ using a mixed-signal capacitive multiplier [12] as follows:

$$\mathbf{y}_{m,i} = \rho_0(x_{max}\mathbf{1} - \mathbf{x}_i) * \mathbf{w}_i + \rho_1 \mathbf{x}_i + \rho_2 \mathbf{w}_i + \boldsymbol{\eta}_{m,i} \quad (7)$$

where $*$ represents element-wise product of two vectors, and $\rho_0$~$\rho_2$ captures the non-linearity due to charge sharing based computation, and $\boldsymbol{\eta}_{m,i}$ is a vector of samples from $\mathcal{N}(0, \sigma_m)$ representing the impact of reset mismatches. The derivation of this model is provided in the supplementary material.

The CBP uses charge sharing-based circuits to sum up the elements of $\mathbf{y}_{m,i}$ and obtain the dot product $\mathbf{w}_i^T \mathbf{x}_i$:

$$y_{s,i} = \mathbf{1}^T \mathbf{y}_{m,i} \quad (8)$$

The residual digital processor maintains a running sum of the row-wise dot products in order to compute the output $y_o = \sum_i y_{s,i} - b$ in (5) followed by $\hat{y} = sign(y_o)$ as the computed label. Equations (6)-(8) describes the behavior of the Compute Sensor. Table 1 lists the model parameters values in a $65nm$ CMOS process. These equations can be employed to estimate the system behavior of the Compute Sensor.

The Compute Sensor's energy consumption per decision, i.e., in processing one $M_r \times M_c$ image, is given by:

$$E_{CS} = M_r M_c (E_p + E_m) + M_r (2E_{adc} + 2E_{add}) + E_{add} \quad (9)$$

where $E_p$, $E_m$, $E_{adc}$, and $E_{add}$ are the energy consumptions of the pixel, capacitive multiplier, the ADC, and a digital adder, respectively.

The conventional system needs to convert all pixel values into the digital domain then process digitally. The energy consumption per decision is given by:

$$E_{conv} = M_c M_r (E_p + E_{adc} + E_{rd}) + M_c M_r E_{mac} \quad (10)$$

where $E_{rd}$ and $E_{mac}$ are the energy per read-out and multiply-accumulate (MAC) operation, respectively. The behavioral and energy model will be employed in Section 4 to evaluate the system performance and energy savings. The energy savings from Compute Sensor are evident from (9) and (10). The key savings arise from having the ADC operate on row-wise dot products giving rise to the multiplicative factor of $2M_r$ as compared to the factor of $M_c M_r$ ($M_c >> 2$) for the conventional system. The second source of energy savings arises from the analog domain multiplication in the Compute Sensor compared to digital domain because $E_{mac} \approx 3E_m$ or $4E_m$.

# 4 Simulation Results

We first validate the behavioral and energy models described in Section 3 using the parameters of a $65\,\mathrm{nm}$ CMOS process. The system performance and energy savings achieved by Compute Sensor are estimated using these models. In the following, the conventional system is assumed to be operating with noise-free data and ideal digital computations.

The Compute Sensor architecture in this study consists of a $32 \times 32$ APS array, a capacitive multiplier array with $5b$ weight, a $10\,\mathrm{b}$ column ADC array, and $16\,\mathrm{b}$ addition in the digital domain. The conventional digital implementation has an identical APS array and ADC, but employs a digital MAC with $10\,\mathrm{b}$ input, $5\,\mathrm{b}$ weight, and $32\,\mathrm{b}$ output. These precisions are the minimum needed for the conventional architecture to achieve a classification accuracy $p_c = 95\%$. The face and non-face images extracted from Caltech101 dataset [8] consisting of $32 \times 32$ gray-scale images is employed.

## 4.1 Model Validation

Table 1 lists the parameters values for the behavioral model in (6) obtained by curve fitting to the results of circuit simulations of a standard 3-transistor APS. The model is found to match detailed circuit simulations to within $5.2\%$ when the pixel output $x_{i,j}$ lies in the interval $[0.2, 0.9]$. The standard deviation of spatial mismatch $\sigma_s$ was found to lie in the interval $[1.62 \times 10^{-2}, 2 \times 10^{-2}]$ using Monte Carlo circuit simulations. The standard deviation of output referred noise $\sigma_n$ was found to lie in the interval $[7 \times 10^{-4}, 7.5 \times 10^{-4}]$. The model parameters in (7) were obtained using the methodology in [12] and are also listed in Table 1.

Table 1: Model parameters in 65nm CMOS

| $x_{max}(V)$ | $\gamma(V/(lx \cdot s))$ | $\sigma_s(V)$ | $\sigma_n(V)$ |
|---|---|---|---|
| 0.9 | $4.39 \times 10^{-5}$ | $2 \times 10^{-2}$ | $7.5 \times 10^{-4}$ |
| $\rho_0$ | $\rho_1$ | $\rho_2(V)$ | $\sigma_m(V)$ |
| 0.93 | $1.2 \times 10^{-2}$ | $6.68 \times 10^{-4}$ | $1.6 \times 10^{-2}$ |

## 4.2 Classification Accuracy

The classification accuracy of Compute Sensor is evaluated employing the models in Section 4.1 with parameters from Table 1. Figure 3(a) shows that the Compute Sensor is able to achieve a classification accuracy $p_c = 94.7\%$ at the nominal values of spatial mismatch $\sigma_s = 2 \times 10^{-2}$, multiplier mismatch $\sigma_m = 2 \times 10^{-2}$, and noise $\sigma_n = 7.5 \times 10^{-4}$. This accuracy which is very close to the value of $95\%$ achieved by the ideal digital implementation. In fact, the Compute Sensor is able to maintain $p_c \geq 94\%$ when $\sigma_s$ is increased to $0.1$, which is $5\times$ more than the nominal value, using the hyperparameters obtained with the nominal value of $\sigma_s$. Any further increases in $\sigma_s$ leads to a large reduction in $p_c$. For example, $p_c$ decreases to $87\%$ when $\sigma_s$ increases to $0.5$. Next, we retrain the Compute Sensor with data generated in the presence of spatial mismatch. Figure 3(a) shows that the Compute Sensor achieves a $p_c = 92\%$ when $\sigma_s = 0.5$ after retraining. This clearly indicates the effectiveness of retraining in order to compensate for spatial mismatch in the APS array. Retraining can also be employed to address the computational errors due to capacitive multiplier mismatch $\eta_m$. A similar study was conducted to observe the impact of multiplier mismatch $\sigma_m$ as shown in Figure 3(b) where $\sigma_s$ and $\sigma_m$ were set at their nominal values. This figure shows that the Compute Sensor achieves $p_c = 90\%$ in the presence of $\sigma_m = 0.5$ with retraining which is a significant improvement over case when retraining was not employed.

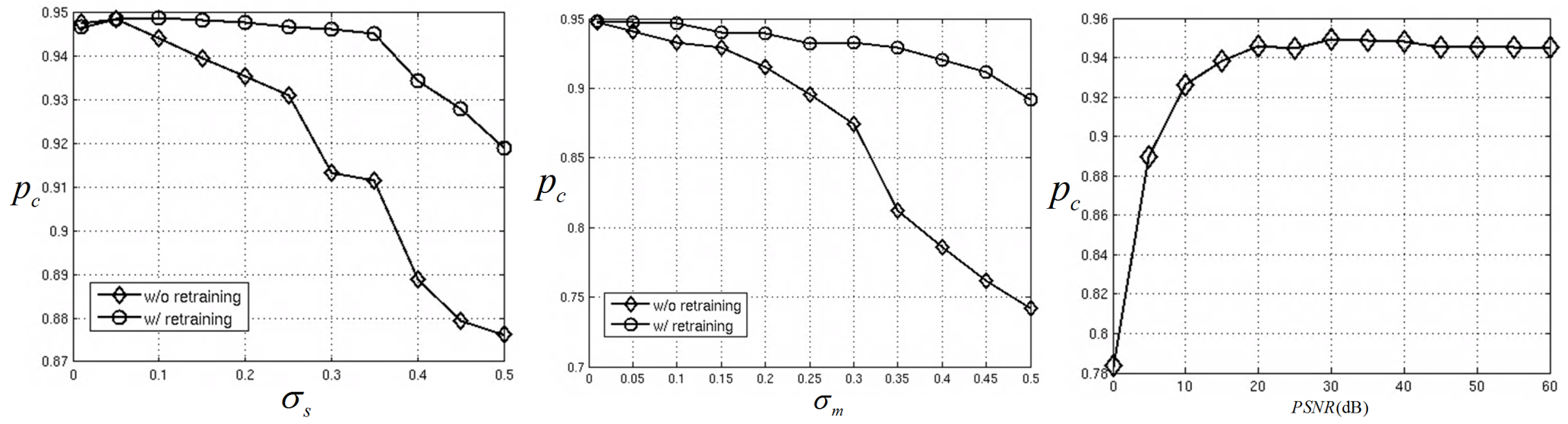

(a) (b) (c)

Figure 3: Classification accuracy of the Compute Sensor wrt.: a) APS spatial mismatch, b) capacitive multiplier mismatch, and c) input peak signal-to-noise ratio ($PSNR$).

Classification accuracy is a function of the input peak signal-to-noise ratio $PSNR$ defined as $PSNR = 20\log_{10}\frac{x_{max}}{\sigma_n}$. A $PSNR = 61\,\text{dB}$ is obtained with the nominal values of $x_{max} = 0.9$, $\sigma_s$, $\sigma_n$, and $\sigma_n$, a $PSNR = 61\,\text{dB}$ is obtained. At this value of $PSNR$, a classification accuracy of $94.7\%$ is achieved. Figure 3(c) shows that the Compute Sensor's classification accuracy decreases to $78\%$ as the $PSNR$ reduces to $0\,\text{dB}$.

Figure 4(a) shows the distribution of the feature vectors when circuit non-idealities are absent. In this case, the SVM chooses a hyperplane that successfully separates the two classes. However, in the presence of spatial and multiplier mismatch ($\sigma_s = \sigma_m = 0.3$), the feature vectors shift as shown in Figure 4(b). The classification accuracy falls if the original hyperplane is used. However, retraining with the new set of feature vectors enables the Compute Sensor to obtain a new separating hyperplane with a commensurate improvement in the classification accuracy as shown in Figure 4(c). These results indicate that the Compute Sensor may need to adapt to changing environmental conditions such as temperature in order to ensure that the optimal separating hyperplane is generated and employed for classification.

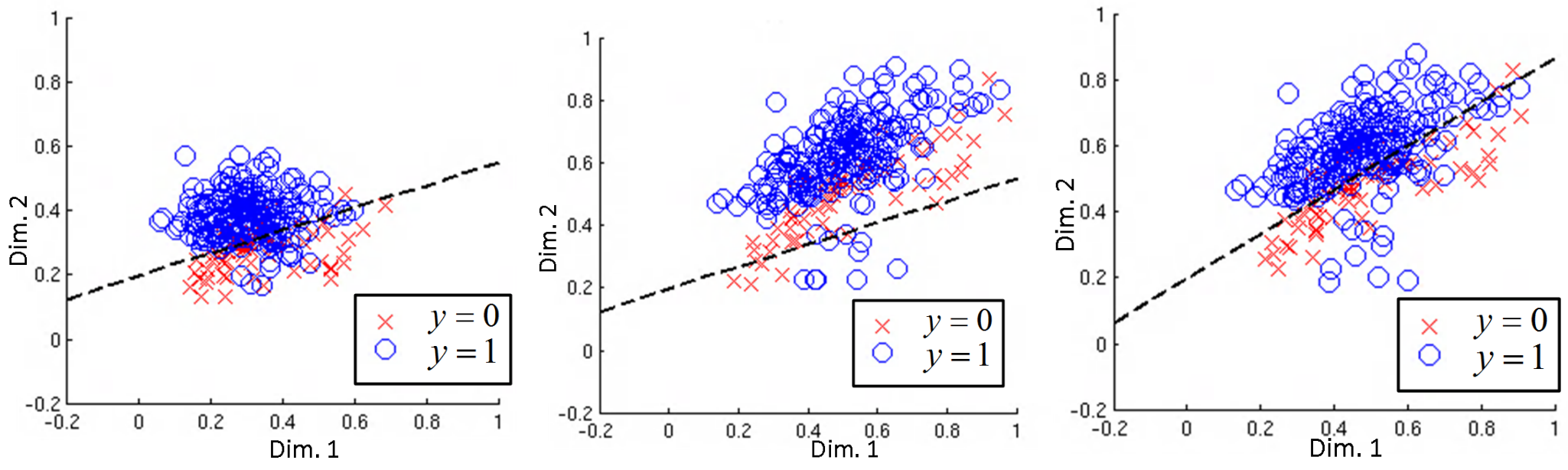

(a) (b) (c)

Figure 4: The feature distribution and SVM separation hyper-plane when: a) $\sigma_s = \sigma_m = 0$ without retraining, b) $\sigma_s = \sigma_m = 0.3$ without retraining, and c) $\sigma_s = \sigma_m = 0.3$ with retraining.

### 4.3 Energy Savings

In order to compare the energy consumption of the Compute Sensor with the conventional architecture, we employ the energy numbers in Table 2 which are based on circuit simulation and published energy numbers from [4, 11].

Table 2: Energy per pixel processing in 65nm CMOS

| $E_p(pJ)$ | $E_{adc}(pJ)$ | $E_{rd}(pJ)$ |
|---|---|---|
| 2.69 | 20.5 | 5 |

| $E_m(pJ)$ | $E_{mac}(pJ)$ | $E_{add}(pJ)$ |
|---|---|---|
| 0.77 | 3.2 | 0.1 |

Figure 5(a) shows that the proposed Compute Sensor consumes $6.2\times$ less energy compared with conventional implementation. The main source of energy savings is due to the elimination of the per bit-line ADC and RD energy and from the use of analog dot product computations in the Compute Sensor. For example, the 1024-length dot product in analog consumes $0.79\,\text{nJ}$, which is $4.1\times$ less than the $3.28\,\text{nJ}$ needed by the digital implementation.

We also studied the energy savings as a function of the array size as shown in Figure 5(b). Indeed, the energy savings increases from $6.2\times$ to $11\times$ as the APS array size increases from $32 \times 32$ to $512 \times 512$. This is because the Compute Sensor performs ADC operations row-wise on dot products, as compared to the conventional architecture which performs ADC operation pixel-wise on scalars.

Another opportunity to reduce energy consumption is to reduce the APS current. However, doing so will degrade the input $PSNR$, as shown in the supplementary material. This may be acceptable if retraining is employed as suggested in Figure 3(c). This figure shows that the Compute Sensor is able to achieve less than $1\%$ performance drop from the ideal digital performance of $p_c = 95\%$ for $PSNR \geq 20\,\text{dB}$. This relaxed $PSNR$ requirement allows the APS array current to be reduced for additional energy savings. Figure 5(c) shows that the energy savings increases to $17\times$ as the $PSNR$ decreases from the $61\,\text{dB}$ to $20\,\text{dB}$.

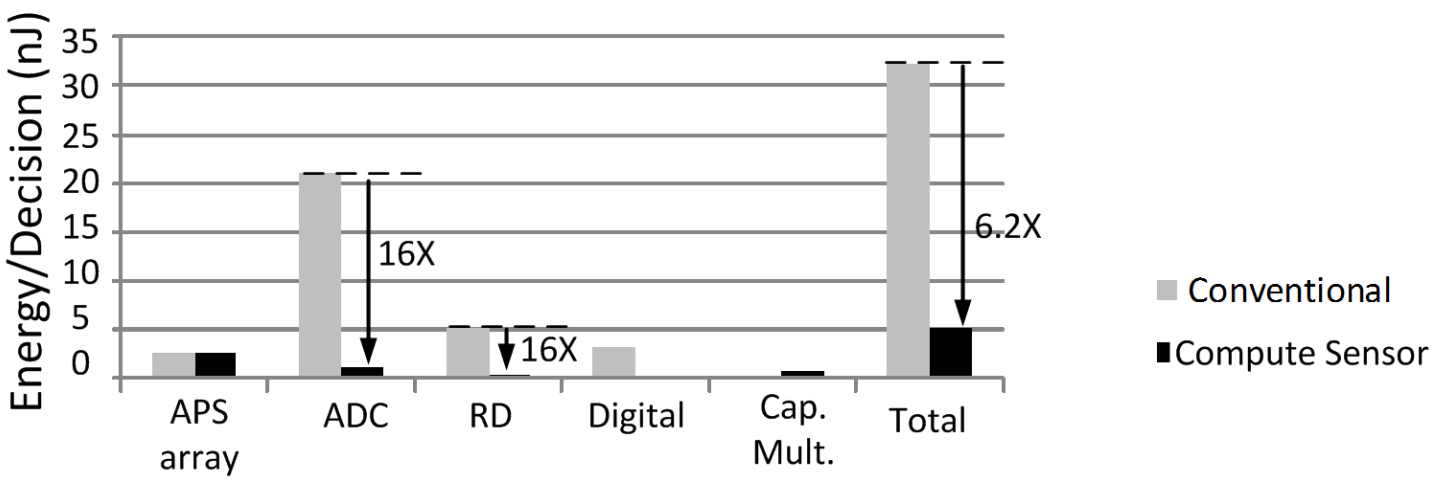

(a)

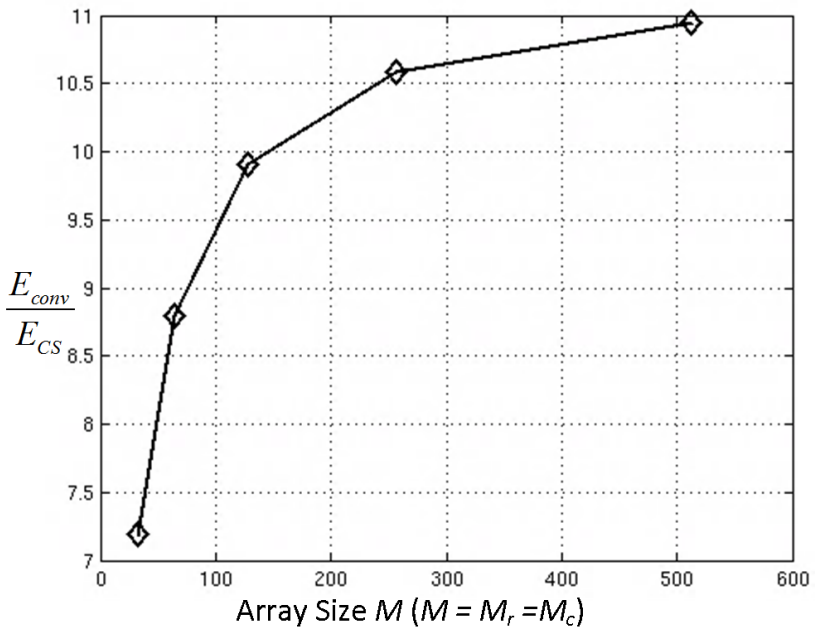

(b)

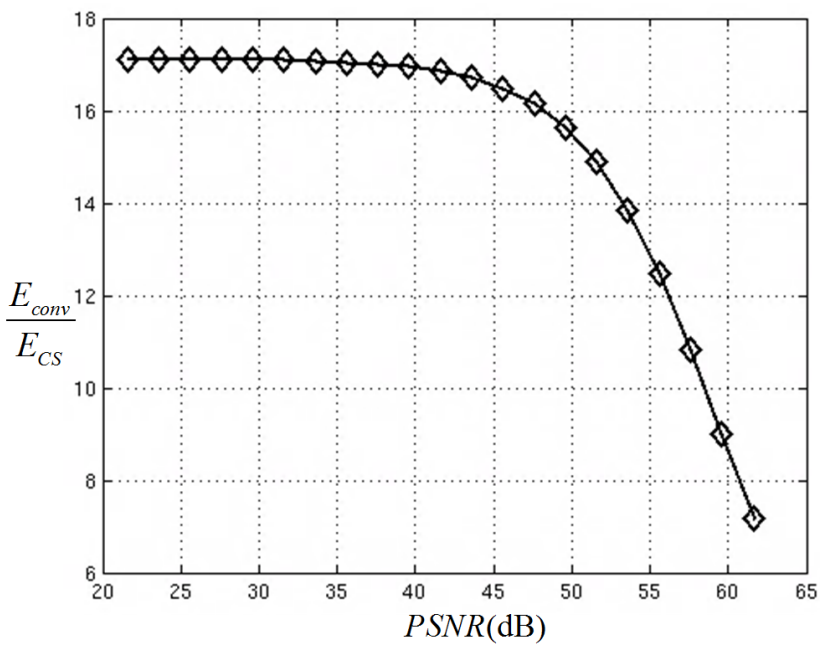

(c)

Figure 5: Energy per decision: a) energy break-down, b) energy savings vs. APS size, and c) energy savings vs. $PSNR$.

## 5 Discussion

We have shown the benefits of embedding information processing functionality into the sensory substrates. We note that such embeddings are made possible due to the intrinsic ability of machine learning algorithms to adapt to noise. Behavioral models such as those in Section 3.2 can be employed to develop a variety of machine learning algorithms for Compute Sensor style architectures. We believe that more powerful machine learning algorithms including deep neural networks, ensemble methods such as bagging and boosting, decision trees, and random forest, can also potentially be embedded into the Compute Sensor. The huge design space spanned by the Compute Sensor encompassing algorithms, architectures, circuits, and sensors, can be a challenge when searching for energy-optimal implementations. Another formidable challenge that we hope to address in the future is to design programmable Compute Sensor architectures whereby a variety of algorithms can be mapped on to the same platform.

## Acknowledgment

This work was supported in part by Systems on Nanoscale Information fabriCs (SONIC), one of the six SRC STARnet Centers, sponsored by MARCO and DARPA.

# Supplementary Material

## 1 Active Pixel Sensor Fundamentals

This section gives background on the circuit operation of active pixel sensor (APS). Solid state imaging devices can be classified into two categories, i.e., charge coupled device (CCD) sensor and CMOS sensor. CMOS image sensor has gain much popularity due to the low voltage, low power operation, and compatibility with standard CMOS technologies APS is by far the most widely employed CMOS image sensor architecture due to the speed advantage and low noise. The architecture of the 3-transistor (3T) APS and associated read-out (RD) circuit is shown in Fig.1(a) where each APS consists of a photodiode (PD) as the optical detector and three transistor for read out. A rolling shutter operation where the pixel array is exposed row by row is typically employed for 3T-APS array, and the timing diagram is shown in Fig. 1(b). During the reset phase, $M_{RST}$ is on and the charge integrated on the photodiode is removed. During the integration phase, $M_{RST}$ is off and the photodiode converts light into current, discharging the parasitic capacitor $C_{PD}$. During the readout phase, $M_{SEL}$ is on, and the signal voltage $V_{SIG}$ is sampled to sampler S-SIG.

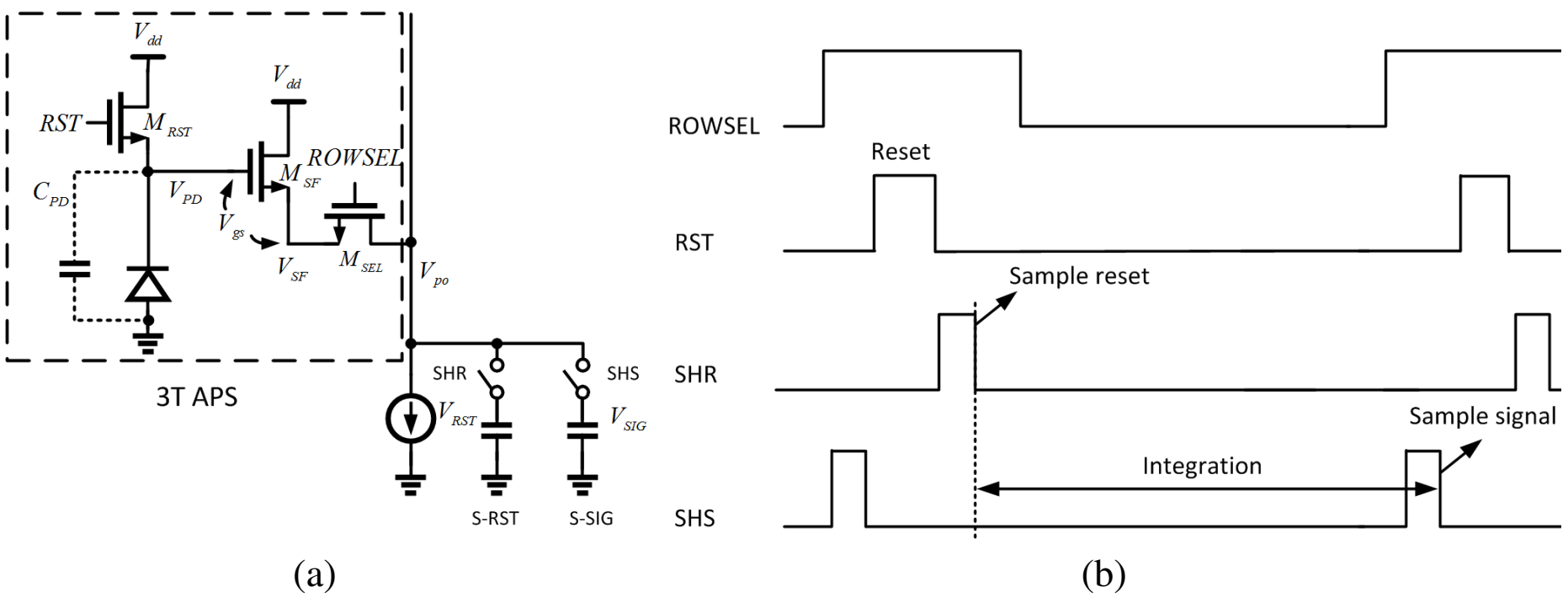

Figure 1: 3T APS: a) pixel architecture and associated RD circuit, and b) timing diagram in rolling shutter operation.

## 2 Compute Sensor Behavioral Model

This section illustrates derivation of the Compute Sensor behavioral model. We focus on the APS and the scaling block.

### 2.1 APS behavioral model

During the APS operation, the exposed PD voltage is first sampled on the sampler in S&H block in Fig. 2(a) by selecting the associated word line (WL). When the $i^{th}$ row is selected, the voltage on the $j^{th}$ sampler $V_{SIG}$ can be expressed as:

$$V_{SIG,j} = V_{PDrst} - V_{gs0} - [\frac{\kappa_1}{C_{PD}} - \kappa_2(V_{gs0} - V_{gs1})]I_{i,j} + \Delta V_{th,j} + V_{n,j} \tag{S.1}$$

where $V_{PDrst}$ is the voltage of the PD after reset, $V_{gs0}$ and $V_{gs1}$ are the gate to source voltage of $M_{SF}$ (see Fig. 1(a)) in dark and highest illumination condition, $C_{PD}$ is the parasitic capacitance at the PD node, $\Delta V_{th,j}$ is the threshold mismatch, $V_{n,j}$ is the output referred RD noise, and $\kappa_1$ and $\kappa_2$ are fitting parameters. The model in (7) can be derived by noting that:

$$x_{max} = V_{PDrst} - V_{gs0} \tag{S.2}$$

$$\gamma = [\frac{\kappa_1}{C_{PD}} - \kappa_2(V_{gs0} - V_{gs1})] \tag{S.3}$$

$$\tag{S.4}$$

and the threshold mismatch $\Delta V_{th,j}$ and noise $V_{n,j}$ are modeled as normally distributed random variables with variances $\sigma_s^2$ and $\sigma_n^2$, respectively.

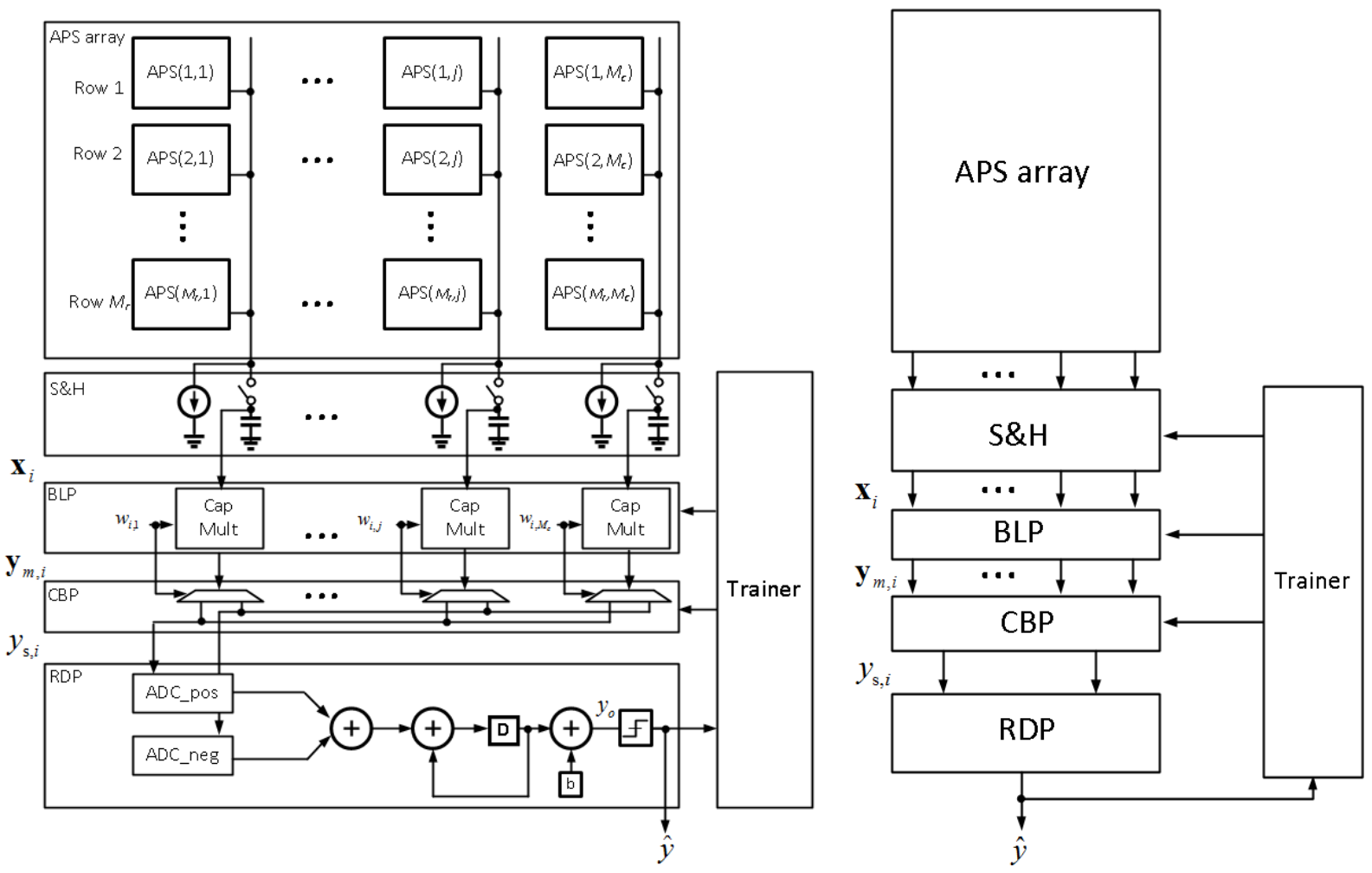

(a) (b)

Figure 2: Compute Sensor-based PCA and SVM: a) architecture, and b) behavioral model.

### 2.2 Scaling block behavioral model

In the Compute Sensor, scaling operation between input $\Delta V_{SIG,i,j} = V_{PDrst} - V_{gs0} - V_{SIG,i,j}$ and weight $w_{i,j}$ is realized in the bit-line processor employing a mixed-signal capacitive multiplier as shown in Fig. 3(a). We next drop the index $(i, j)$ and denote the analog voltage and $B_p$-b digital weight as $\Delta V_{SIG}$ and $w = \sum_{i=0}^{B_p-1} p_i 2^{-(B_p-i)}$, respectively, for notational simplicity. The capacitive multiplier employs successive charge sharing to obtain a voltage $V_m$ of:

$$V_m = V_{pre} - w(V_{pre} - (V_{PDrst} - V_{gs0}) - \Delta V_{SIG}) \tag{S.5}$$

By choosing $V_{pre} = V_{PDrst} - V_{gs0}$, the voltage drop $\Delta V_m$ is thus:

$$\Delta V_m = V_{pre} - V_m = w\Delta V_{SIG} \tag{S.6}$$

To account for the nonlinearity due to charge sharing based operation and the mismatch in the reset transistors, the following model is employed:

$$\Delta V_m = \rho_0 \Delta V_{SIG} w + \rho_1 V_{SIG} + \rho_2 w + \eta_m \tag{S.7}$$

The behavior in (8) can be obtained by noting that $y_m = \Delta V_m$ and $\Delta V_{SIG} = V_{PDrst} - V_{gs0} - V_{SIG} = x_{max} - x$.

## 3 Energy $PSNR$ trade-off

One of the fundamental noise sources in APS is the thermal noise, whose noise power is described by:

$$\sigma_n^2 = kT/C \tag{S.8}$$

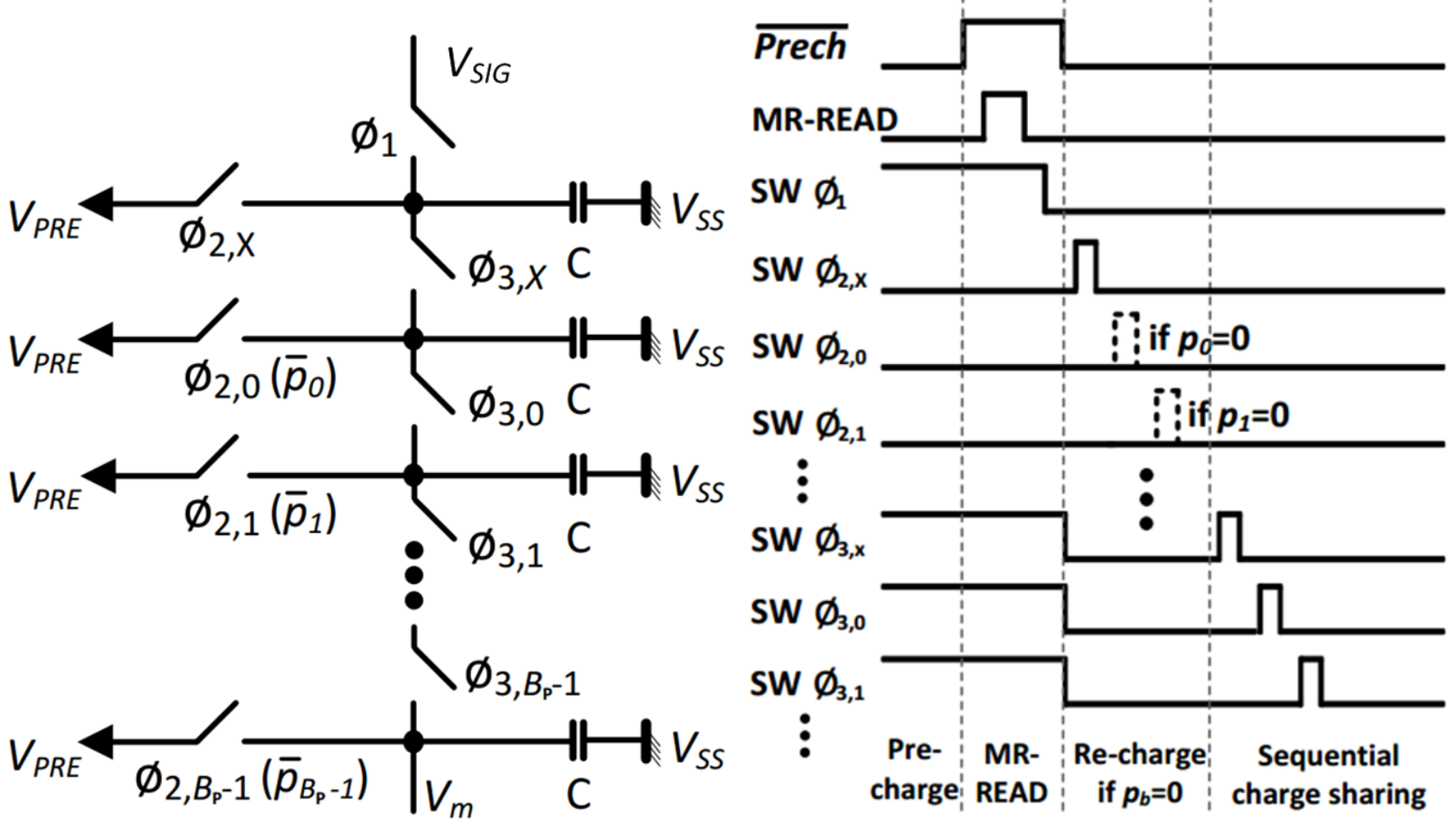

(a) (b)

Figure 3: Capacitive multiplier a) architecture and b) timing diagram.

where $k$ is the Boltzmann's constant, $T$ is the temperature, and $C$ is the sampling capacitance. The bandwidth of the APS can be approximated by:

$$B = \frac{g_m}{C} = \frac{I_{aps}}{V_{ov}C} \tag{S.9}$$

where $I_{aps}$ is the current consumption of the APS array, $V_{ov}$ is the overdrive voltage and is a fixed parameter chosen during the design. A fundamental trade-off between noise and bandwidth (thus speed) can be seen from (S.8) and (S.9). For fixed bandwidth, reducing $C$ will allow smaller $I_{aps}$ thus lower energy, but will increase the noise variance $\sigma_n^2$ per (S.8). More specifically, the $PSNR$ is related to the current $I_{aps}$ via:

$$PSNR = 20log_{10}(x_{max}/\sigma_n) \propto 10log_{10}(I_{aps}) \tag{S.10}$$

and the energy of the APS is related with $I_{aps}$ via:

$$E_{pix} = V_{dd}I_{aps}T_{pix} \tag{S.11}$$

where $V_{dd}$ and $T_{pix}$ is the supply voltage and the pixel access time, respectively. Equation (S.10) - (S.11) is employed to obtain Fig. 5(c).